# A Rhetorical Relations-Based Framework for Tailored Multimedia Document Summarization

[1]Azze-Eddine Maredj, [2]Madjid Sadallah

[1,2,3] DTISI, Research Center on Scientific and Technical Information CERIST, 05 Rue des 3 Frères Aissou, 16028, Ben Aknoun, Algiers, Algerie

[1] amaredj@cerist.dz, [2] msadallah@cerist.dz

*Abstract*—*In the rapidly evolving landscape of digital content, the task of summarizing multimedia documents, which encompass textual, visual, and auditory elements, presents intricate challenges. These challenges include extracting pertinent information from diverse formats, maintaining the structural integrity and semantic coherence of the original content, and generating concise yet informative summaries. This paper introduces a novel framework for multimedia document summarization that capitalizes on the inherent structure of the document to craft coherent and succinct summaries. Central to this framework is the incorporation of a rhetorical structure for structural analysis, augmented by a graph-based representation to facilitate the extraction of pivotal information. Weighting algorithms are employed to assign significance values to document units, thereby enabling effective ranking and selection of relevant content. Furthermore, the framework is designed to accommodate user preferences and time constraints, ensuring the production of personalized and contextually relevant summaries. The summarization process is elaborately delineated, encompassing document specification, graph construction, unit weighting, and summary extraction, supported by illustrative examples and algorithmic elucidation. This proposed framework represents a significant advancement in automatic summarization, with broad potential applications across multimedia document processing, promising transformative impacts in the field.*

## I. INTRODUCTION

In the contemporary digital landscape, the proliferation of multimedia content, encompassing text, images, and videos, presents a significant challenge: how can we efficiently summarize and comprehend these heterogeneous documents [1]? Automatic document summarization has emerged as a promising approach, aiming to distill essential information and generate succinct summaries that encapsulate the core of the original content [2]. Summarizing multimedia documents, characterized by their multifaceted content types, necessitates a system capable of accommodating this diversity while ensuring that the resultant summary is both succinct and coherent, faithfully preserving the meaning of the original document.

Integral to multimedia document summarization is the pivotal task of evaluating the importance of each document entity, spanning text, image, or video. This necessitates a thorough assessment of the significance of each component, determining which entities warrant inclusion in the summary. Additionally, maintaining coherence presents another challenge, as the summary must not only underscore essential entities but also present them in a manner that preserves the original meaning and structure of the document.

To tackle these challenges, researchers have proposed various methodologies, among which is the utilization of Rhetorical Structure Theory (RST). Originally formulated by Mann and Thompson [3], RST provides a robust framework for analyzing and representing coherence relations within textual documents. It elucidates how different segments of a document are interconnected and contribute to its overall meaning. In our prior research, we introduced a media weighting approach that extends the conventional rhetorical structure model [4]. This extension incorporates rhetorical relations between diverse media types, thereby enhancing the modeling of multimedia documents. By leveraging RST, this approach facilitates the determination of importance, rendering it applicable to tasks such as document adaptation, automatic composition, and summary generation. Expanding upon this groundwork, our present study introduces an innovative framework for multimedia document summarization. Harnessing RST, our framework analyzes the document structure to ascertain the significance of each entity, ensuring the production of summaries that are both succinct and faithful to the original content.

The subsequent sections of the paper are organized as follows: Section 2 reviews related work in document summarization and RST utilization. Section 3 details our proposed framework, elucidating the steps involved in entity importance determination and coherence preservation. Section 4 outlines the experimental setup and evaluation metrics used to assess our framework's performance. Finally, Section 5 concludes the paper, summarizing findings and discussing future directions.

## II. BACKGROUND & RELATED WORKS

### A. Structural and Semantic Document Modeling

Effectively representing the intricate structures inherent in multimedia documents is crucial for various tasks such as analysis, retrieval, summarization, generation, and adaptation. Traditional models, like the four-dimensional model [5], encapsulating temporal, spatial, logical, and hypermedia dimensions, have limitations in addressing media integration, adaptation, and personalized consumption. For instance, the four-dimensional model widely used fails to comprehensively address the challenges posed by the dynamic nature of multimedia content.

To overcome these limitations, researchers have introduced extensions and alternative models. André et al. [6]

presented a five-dimensional model, enriching it by incorporating user interaction as a fundamental dimension. In [7], this model was leveraged to assist reengineering documents based on readers' usages. Nack and Hardman [8] emphasized the importance of semantics in document modeling, enabling a more profound capture of meanings embedded within multimedia documents. The exploration into adaptive hypermedia, as seen in the work of Brusilovsky et al. [9], has delved into the customization of content based on user characteristics, adding a layer of personalization to the traditional models. In [10], the authors proposed the addition of an annotation dimension to the four model.

Geurts et al. [11] proposed an ontology-based model, utilizing ontologies to represent domain knowledge, document content, and user profiles. This innovative approach provides a more nuanced understanding of the intricate relationships within multimedia documents. Additionally, Yao et al. [12] introduced a graph-based model that goes beyond structural representation and illustrates both the structural and semantic relationships between various components of multimedia content.

In this context, the concept of rhetorical structure, encompassing the organization and arrangement of media elements, has emerged as a critical consideration in multimedia design, significantly influencing the clarity, effectiveness, and overall appeal of a document. Despite the diversity of document forms, researchers have dedicated efforts to develop frameworks addressing this intricate aspect. Haase's [13] analysis of narrative structures in interactive multimedia applications, exploring narrative media and narrativity, contributes valuable insights into the nuanced interplay inherent in narrative design. In this context, Rhetorical Structure Theory (RST), introduced by Mann and Thompson [3], aligns seamlessly with the broader concept of rhetorical structure. Serving as a popular discourse structure theory, RST utilizes a tree structure to represent text coherence, making it particularly well-suited for high-level tasks such as text summarization.

*B. Graph-based Representations of Documents*

While traditional processing methods, such as the bag-of-words (BoW) model, have been widely used, they may exhibit limitations in their efficacy. The BoW model treats documents as unordered collections of terms, neglecting their sequence and relationships, thereby restricting its utility in tasks requiring an understanding of a document's structure and meaning. In response to these limitations, graph-based representations of documents emerge as a promising solution.

By constructing graphs that illustrate connections between terms, these representations capture both semantic nuances and structural elements of textual data [14], enhancing various natural language processing tasks like text classification and information retrieval. However, creating meaningful graphs from raw documents presents challenges, prompting researchers to employ strategies such as syntactic parsing, semantic analysis, and topic modeling to augment graph structures with valuable content for downstream applications.

Graph-based representations have proven effective in various natural language processing tasks. For instance, Fei et al. [15] introduced a span-graph neural model for biomedical texts, capturing intricate relationships between terms. Another study by Fei et al. [16] addressed semantic role labeling using a label-aware graph convolutional network (LA-GCN) to encode syntactic arcs and labels into BERT-based word representations, capturing rich semantic and syntactic information.

In social network analysis, Khanday et al. [17] demonstrated the effectiveness of graph-based representations in detecting propaganda and propagandistic communities. Graph-based representations extend to multimedia documents, as shown by Hochin and Nishida [18], who introduced a model capturing semantic relationships between text and images. In multimedia recommendation, Zhang et al. [19] proposed techniques for learning item-item structures across modalities, and Wei et al. [19] suggested a hierarchical user intent graph network.

*C. Multimedia Document Summarization*

The field of multimedia document summarization has seen significant advancements over the years due to the exponential growth of multimedia content [8]. The task involves extracting key information from multimedia documents, which comprise text, images, and videos. The challenge lies in effectively summarizing these diverse types of content while maintaining the coherence and relevance of the summary.

Graph-based techniques have been extensively employed in multimedia document summarization. In [8], two methods for summarizing multimedia content are proposed. Their approach involved creating a graph representation of the document and using weighting algorithms to assign importance values to the nodes, which represent different units of content. However, this approach does not consider the inherent structure and coherence of the document, which can lead to summaries that lack context and relevance.

In addition to graph-based techniques, the Semantic Adaptation Framework has been proposed as a method for dealing with the semantics of document composition by transforming the relations between multimedia objects [3]. This technique also has the capability of suppressing multimedia objects, which is useful for summarizing multimedia documents.

Recent advancements in the field have seen the introduction of multimodal approaches to document summarization. Qiu et al. [21] proposed the Multimodal Hierarchical Multimedia Summarization (MHMS) framework which interacts with visual and language domains to generate both video and textual summaries. While this represents a significant advancement in the field, the MHMS framework does not consider user preferences and desired summary time, which can limit the personalization and contextual relevance of the summaries.

Semantic graphs for web document visualization have also been explored [4]. These graphs provide a visual representation of the semantic relationships between different units of content, facilitating the understanding and navigation of complex multimedia documents. However, these semantic graphs do not assign importance values to the units of content, which can result in summaries that do not accurately represent the original document.

Deep learning models have also been used for multi-document summarization [22]. These models can learn complex patterns and relationships in the data, making them effective for this task. For instance, a recent work proposed a multimodal hierarchical multimedia summarization (MHMS) framework that interacts with visual and language domains to generate both video and textual summaries documents [2]. Multi-document summarization involves generating a summary from multiple. This concept is particularly relevant to multimedia document summarization, as it involves dealing with large volumes of diverse content. However, multi-document summarization techniques often struggle with maintaining the coherence of the summary when dealing with large volumes of diverse content.

*D. Identified Gaps and Rationale for the Proposed Framework*

While significant progress characterizes multimedia document summarization, scrutiny uncovers noteworthy gaps in existing methodologies. These limitations underscore the need for a refined approach, motivating the development of our framework. Graph-based approaches, exemplified in [8], excel in assigning importance values to individual nodes but often struggle to maintain the inherent coherence and structure of the entire document. This challenge becomes pronounced when aiming for comprehensive summaries that encapsulate the document's meaning and context. In multimodal approaches, like the MHMS framework [21], advancements in handling diverse content types are evident. However, a lack of personalized summarization persists, as these approaches overlook user preferences and desired time frames, limiting adaptability to individual needs. Current models, including the Semantic Adaptation Framework [3], may not fully exploit the inherent structure of multimedia documents during summarization, resulting in summaries lacking cohesiveness and context.

In response, our framework for multimedia document summarization is introduced. It aims to overcome existing challenges by emphasizing document coherence, leveraging Rhetorical Structure Theory (RST) to capture rhetorical relations between different media types. The framework goes beyond conventional graph-based representations by integrating user preferences and desired summary time into weighting algorithms, enabling a personalized and contextually relevant summarization experience.

## III. A FRAMEWORK FOR DOCUMENT SUMMARIZATION

Summarizing multimedia documents is a challenging task due to the diverse media types and intricate content relationships. Our framework takes a systematic approach, blending theory with practical methodologies. Here's an overview of key steps and their rationale (Fig. 1):

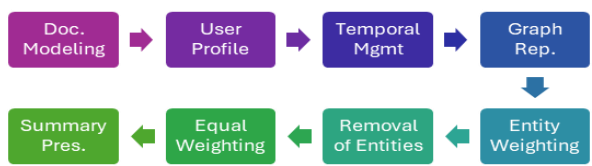

Fig. 1 Representation of the Summarization Process

1. *Document Modeling* uses rhetorical structure modeling to understand multimedia documents.
2. *User Profile Integration* aligns summaries with user's preferences and interests.
3. *Temporal Management* controls summary length and duration based on user's time constraints.
4. *Graph-based Representation* uses a rhetorical relations graph to depict relationships and hierarchies.
5. *Entity Weighting* assigns weights to entities based on relationships, unit types, and structures.
6. *Removal of Entities* systematically removes less important entities within specified constraints.
7. *Equal Weighting Handling* prioritizes entities based on user preferences and rhetorical roles in cases of equal weighting.
8. *Summary Presentation* displays the summary in various formats adapted for different devices.

Each of these stages contributes to the generation of concise and informative summaries that are tailored to user preferences and the desired summary duration. This approach ensures a systematic and comprehensive process for multimedia document summarization.

*A. Document Modeling and Specification*

*1) Rhetorical Structure Modeling*

Rhetorical Structure Theory (RST) provides a powerful foundation for understanding and representing the coherence relationships within textual content. In the context of multimedia document summarization, we capitalize on the inherent capabilities of RST to enhance the representation and analysis of complex structures encompassing diverse media types. At its core, RST captures authorial writing styles and intent by representing a document as a tree-like structure. This structure comprises leaves as Elementary Discourse Units (EDUs) and internal nodes as contiguous text spans with rhetorical relations. EDUs are coherent textual units, ranging from simple sentences to complex clauses, each possessing independent syntactic and semantic meaning. The construction of the RST tree structure involves the segmentation of text into EDUs, followed by the sequential merging of adjacent units based on the presence of rhetorical relations. These relations encapsulate the interconnections between two non-overlapping text spans, resulting in a cohesive whole. Each text span is designated as a nucleus or satellite based on the relation.

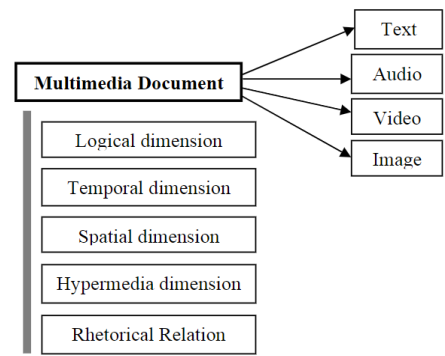

Fig 2: Enriched Document Model with Rhetorical Relations Dimension

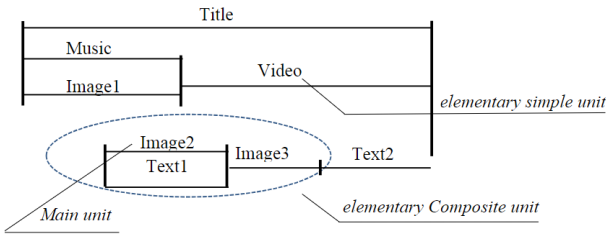

Fig 3. Elementary Composite Unit and Main Unit

In our proposed framework [4], we extend the conventional four-dimensional model of multimedia documents by incorporating a vital dimension focused on rhetorical relations among various media elements. This extension enriches the discourse structure conveyed by diverse media types, providing a more comprehensive representation, as illustrated in Fig. 2.

Our extension introduces a structured framework based on extended orbits, allowing for the development of algorithms to evaluate the importance of each medium within the document. This innovative dimension significantly influences document analysis, organization, and retrieval processes. To align with RST, we introduce two distinct unit types:

1. Elementary Simple Unit (ESU): This unit type encompasses textual content, images, videos, or audio elements.
2. Elementary Composite Unit (ECU): This unit type comprises a collection of simple units, with a designated main unit serving as its central component, as illustrated in Fig. 3.

Our approach draws inspiration from traditional textual analysis techniques, extending the concept of an "orbit" to the realm of multimedia documents. Leveraging RST, we model rhetorical relationships between Elementary Semantic Units (ESUs) and Elementary Composite Units (ECUs), as visually represented in Fig. 4.

Inspired by traditional textual analysis techniques, our approach expands the concept of an "orbit" to multimedia documents. Leveraging RST, we model rhetorical relationships between Elementary Semantic Units (ESUs) and Elementary Composite Units (ECUs). By extending the notion of an "orbit" to multimedia documents, we utilize RST to encapsulate intricate rhetorical relations between ESUs and ECUs, effectively capturing both semantic and functional dimensions. This innovative modeling approach provides a robust foundation for understanding and navigating nuanced relationships within multimedia documents, thereby contributing to the effectiveness of subsequent summarization processes.

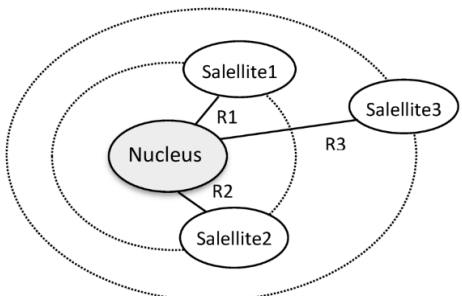

Fig. 4 Orbital Representation of a Nucleus and Its Satellites

### 2) 3.1.2 Bridging Rhetorical Structure and Document Components in Specification

Building upon the foundation laid by Rhetorical Structure Theory (RST) in our framework, the document specification process serves as a pivotal step that aligns with both RST principles and the unique dimensions introduced in the extended model. During this phase, the multimedia document (D) undergoes a comprehensive analysis to ascertain its structure and content, comprising various media types.

The specification process involves identifying distinct units (U = {u1, u2, ..., un}), each representing a coherent piece of information within the document. These units, whether Elementary Simple Units (ESUs) or Elementary Composite Units (ECUs), play a crucial role in contributing to the overall rhetorical relationships within the document.

To align with the principles of Rhetorical Structure Theory, different media types are analyzed using appropriate techniques. For textual content, Natural Language Processing (NLP) techniques can be employed to identify paragraphs, sentences, and finer-grained units. Simultaneously, computer vision algorithms are utilized for images and videos to detect meaningful frames or segments. This dual approach ensures that the document specification process not only considers the text-based rhetorical relationships but also extends its reach to visual and auditory elements. By doing so, our framework encapsulates a more holistic understanding of the multimedia document's discourse, acknowledging the diversity of media types.

Beyond serving as a foundational step for subsequent processes, document specification significantly contributes to the rhetorical structure outlined by RST. The identified units, representing coherent pieces of information, are categorized based on their nature and role within the document. This categorization plays a pivotal role in understanding and enhancing the rhetorical relationships, as each unit contributes uniquely to the document's discourse.

Breaking down the multimedia document into units allows for the capture of important elements and their interconnections. This detailed analysis proves crucial in generating a coherent and informative summary. The resulting units (U = {paragraph1, sentence1, sentence2, image1, video1, video2, ...}) represent not only isolated pieces of information but also interconnected elements that influence the overall understanding of the topic.

*Example*: Consider a multimedia document on "The History of Space Exploration." During document specification, the text is meticulously divided into paragraphs and sentences, while images and videos are segmented into key frames and clips. The resulting units represent distinct pieces of information (U = {paragraph1, sentence1, sentence2, image1, video1, video2, ...}), contributing synergistically to the overall understanding of the topic and influencing the rhetorical structure outlined by RST.

### B. User Profile for Personalizing Summarization

Building upon the document specification process and the enriched rhetorical structure, our framework integrates user profiles into the multimedia document summarization process. This integration aims to personalize summaries according to individual user preferences, interests, and requirements.

The user profile is instrumental in personalizing the summarization, capturing essential information about the user's background, topic preferences, desired level of detail, and areas of interest. This information aligns the summarization output with the user's expectations, ensuring a relevant and engaging summary. The introduction of the user profile involves gathering information through various means such as user surveys, explicit feedback, browsing history, or previous interactions with the summarization system. This approach provides a comprehensive understanding of the user's preferences, enriching the summarization process with personalized insights. Once constructed, the user profile integrates into the summarization process, enabling the system to adapt its output based on the user's preferences, thereby enhancing the relevance and usefulness of the generated summary.

The user profile influences both the content and style of the generated summary. For example, if the user has a background in astrophysics, the system may prioritize detailed information about space exploration milestones and scientific discoveries. If the user prefers concise summaries, the system may focus on delivering key insights while omitting less critical details. The introduction of the user profile transforms summarization into a personalized experience, allowing the system to cater to the specific needs and interests of each user. This adaptive approach results in summaries that are not only more relevant and engaging but also uniquely tailored to individual preferences.

*Example*: Consider a user profile for a space enthusiast with a background in astronomy within the context of *The History of Space Exploration* example. The user profile, P, includes details about the user's educational background, specific interests in space missions and astronomical discoveries, and a preference for detailed summaries. By integrating this user profile into the summarization process, the system can generate summaries that prioritize in-depth coverage of significant space missions, scientific breakthroughs, and historical events related to space exploration. This ensures that the summary aligns with the user's preferences, providing relevant and engaging content tailored to their specific interests.

### C. Temporal Control in Document Summarization

Building upon the personalized summarization enabled by user profiles, our framework introduces an additional layer of user control and convenience—temporal management. Recognizing the crucial role of time in multimedia documents, the introduction of the desired summary time is a key step in controlling the length and duration of the generated summary.

The desired summary time serves as a fundamental constraint, providing users the flexibility to customize the summary length based on their preferences and constraints. This temporal dimension is particularly significant in multimedia document summarization, where diverse content types may have varying information density and duration.

During the summarization process, the desired summary time is a primary consideration, ensuring the generated summary adheres to the specified duration. Advanced algorithms and techniques are utilized to select and prioritize the most relevant and informative entities from the document while maintaining the desired summary time. This step ensures the summary is not only concise but also time-efficient, meeting the user's specific requirements.

The implementation of the desired summary time optimizes both the length and duration of the summary, offering users a convenient and time-saving means to consume essential content from the document. This temporal management allows users to efficiently extract key information without the need to review the entire document, enhancing the accessibility and digestibility of the summary.

*Example*: Consider a user who is interested in *The History of Space Exploration* document and wishes to obtain a summary within a limited timeframe of 10 minutes (T = 10 mn). The summarization system will generate a concise summary encapsulating the most crucial events, missions, and scientific breakthroughs in space exploration—all within the specified 10-minute constraint. This ensures that the summary is not only comprehensive but also time-efficient, providing the user with a thorough understanding of the subject within the desired timeframe.

### D. 3.4 Rhetorical Structure Graph Construction

In the dynamic field of multimedia document summarization, establishing a robust representation of rhetorical structure is foundational for effective summarization. Drawing inspiration from established work [1], our framework adopts a graph-based approach to construct a model featuring nodes, edges, and associated properties. This approach enriches the portrayal of document coherence, aligning with the nuanced relationships highlighted in the rhetorical structure.

- *Nodes*: Representing *Elementary Discourse Units* (EDUs), these nodes encapsulate fundamental units of discourse structure, such as sentences or image captions. They serve as the building blocks for coherent document representation.
- *Edges*: Labeled with types like Cause, Contrast, or Elaboration, these rhetorical relationships connect nodes, creating an interconnected graph that ensures coherence within the document. The identification of these relationships is guided by rhetorical cues and coherence markers.
- *Node Weights*: Reflecting the importance of EDUs in the document context, node weights contribute to prioritizing content during the subsequent summarization steps.

To construct the rhetorical relations graph (G), we use the following algorithm:
1. *Initialize an empty graph G*: Begin with an empty canvas, preparing for the construction process.
2. *Identify document units* (U = {$u_1$, $u_2$, ..., $u_n$}): Referencing the units identified during the document specification step (Section 3.2), this step captures various media types, including sentences, images, and videos.
3. *Determine relationships $R(u_i)$ for each unit $u_i$*: Guided by rhetorical cues and coherence markers, relationships between units are discerned, forming the basis for the graph structure.
4. *Add nodes to graph G*: Each unit ui corresponds to a

node, and nodes are systematically added to the graph.
5. *Add edges to graph G*: Rhetorical relationships between units are translated into edges, weaving the interconnected graph that mirrors the document's rhetorical structure.

The graph-based representation is a robust tool for analyzing and visualizing relationships between document units, aiding in pattern identification, cluster detection, and critical unit ranking, particularly useful for large documents. After specifying diverse media types in the document, the rhetorical relations graph G is carefully constructed. Relationships between units, such as paragraphs, sentences, images, and videos, are discerned based on rhetorical cues and coherence markers. For example, a sentence describing a space mission may establish a nucleus-satellite relationship with an image illustrating that mission. This resulting graph provides a visual overview of the document's structure, shedding light on interconnections and rhetorical significance. It enhances understanding of document coherence, serving as a foundation for subsequent summarization steps.

**Example**: Returning to the *The History of Space Exploration* example where various units (U = {paragraph1, sentence1, sentence2, image1, video1, video2, ...}) were identified during document specification, these units are instrumental in constructing the rhetorical relations graph G. This visual representation elucidates relationships between different units, such as the connection between a textual description of a space mission and an image illustrating that mission. The resulting graph provides a clear understanding of the document's structure, facilitating the identification and prioritization of critical units for the subsequent summarization process.

*E. Weighting and Ranking of Entities*

*1) Unit Weighting*

Unit weighting plays a crucial role in determining the relative importance of each unit within a document, enhancing our understanding of relationships among document units, and facilitating unit ranking based on significance. The importance of a unit is assessed considering factors such as its role as a nucleus or satellite, orbital position, and relationships with other units.

The weighting algorithm we proposed in our work traverses the graph, assigning weights based on relationships, unit types (ESU or ECU), and hierarchical structure [4]. The algorithm ensures a nuanced representation, contributing to improved document comprehension and navigation. The algorithm consists of steps outlined in the following algorithm.

Algorithm: Weighting Algorithm
1. Start with the first inserted unit, assigned the value P.
2. If the starting unit is a composite elementary unit, find its main unit and assign it the value P.
3. Traverse the graph of relationships using the following procedure:
1. At each new relationship, check the type of the destination unit:
 - If it is an elementary simple unit:
   - If it is a multi-nucleus relationship, the units linked by the relationship will have the same importance value as the starting unit.
2. If it is a nucleus/satellite relationship:
3. If it is a satellite, its value is calculated based on its orbit (sum of the weights of all incoming edges).
4. If it is a nucleus, increment its value by 1 compared to the value P.
5. If it is a composite elementary unit, find its main unit and continue the weighting with that unit.
4. The weighting algorithm considers the significance of each unit in the document, using it to assign weights to each unit.

*2) 3.5.2 Graph Levels Computation Algorithm*

To comprehend the hierarchical relationships inherent in multimedia documents, our algorithm assigns levels to each unit through a meticulous depth-first traversal. This process is instrumental in unveiling the hierarchical structure of the document, facilitating a clearer representation of relationships between nuclei and satellites.

Algorithm: Graph Levels Computation
- Start with the first inserted unit, assigned level 0.
- For each new relationship:
- If the destination unit is an elementary simple unit, increment its level by 1.
- If it is a composite elementary unit, find its main unit and continue computing levels with that unit.

This algorithm systematically establishes hierarchical levels for document units, ensuring a structured understanding of their relationships within the rhetorical structure. The depth-first traversal allows for a comprehensive exploration of these relationships, contributing to the overall effectiveness of the summarization process.

*Example:* In the illustrative example *The History of Space Exploration*, the rhetorical relations graph is meticulously constructed with an emphasis on relationships and hierarchical structure. The application of the Graph Levels Computation Algorithm helps to unveil the hierarchical organization of units, making it evident how different elements, such as paragraphs, sentences, images, and videos, are interconnected. This hierarchical understanding, coupled with unit weighting, plays a pivotal role in identifying significant units, ultimately contributing to the creation of a nuanced and coherent summary that effectively captures the essence of the multimedia document.

*F. Deletion of Least Important Entities:*

In the pursuit of condensing the summary to fit the desired time frame, our algorithm systematically removes entities of lower importance while considering the relevance of each entity and adhering to the specified summary time. The algorithm's steps are outlined below.

*Algorithm: Deletion of Least Important Entities*

*Input: Weighted and ranked entities, Desired summary time T*

*Output: Trimmed summary entities*

*- Initialize the trimmed summary entities list.*

*- Iterate through the ranked entities in descending order, starting from the least important.*

*- Check if adding the current entity to the trimmed summary exceeds the desired summary time T.*

*- If the entity can be added without exceeding the time*

*constraint, include it in the trimmed summary.*
- *If the entity is a core entity, remove all its associated satellite entities from the trimmed summary.*
- *Continue the iteration until the desired summary time T is reached or all entities have been considered.*
- *Return the trimmed summary entities.*

This algorithm ensures that the final summary remains within the desired time constraint by progressively excluding the least important entities. Through iterative evaluation of the time constraint and the significance of entities, the algorithm constructs a succinct summary that encapsulates the most crucial and relevant information within the specified time limit.

**Example**: In the example *The History of Space Exploration*, assuming a desired summary time of 5 minutes, the algorithm initiates by considering the least important entity from the ranked entities. It evaluates whether adding that entity to the trimmed summary would exceed the 5-minute time constraint. If not, the entity is included. In the case of a core entity, all associated satellite entities are also removed from the trimmed summary. The algorithm iterates through entities until either the desired summary time is reached or all entities have been considered. The resulting trimmed summary entities represent the main content within the specified time limit.

*G. Treatment of Cases of Equal Weighting:*

In situations where entities possess equivalent weights, supplementary strategies are employed to resolve ambiguity and guarantee the production of an optimal summary. The methodology adapts according to the type of media and the roles designated within the rhetorical structure, thereby fostering flexibility and user participation to fine-tune the summarization process.

- *Entities of different media types*: user preferences steer the decision-making process by forming a hierarchy among various media types. The entity linked with the user's favored media type is given priority in the final summary.
- *Entities of the same media type*: the algorithm evaluates the role of each entity, leveraging the partial order derived from rhetorical relations. Entities playing more influential roles, such as expressing main ideas or central concepts, are given precedence.
- *Persisting equal weights*: user intervention becomes mandatory, enabling users to make the ultimate decision based on their individual preferences.

The algorithm addressing cases of equal weighting unfolds as follows:
*Algorithm: Treatment of Cases of Equal Weighting*
*Input: Weighted and ranked entities*
*Output: Finalized summary entities*
- *Iterate through the weighted and ranked entities.*
- *Identify groups of entities with equal weights.*
- *For groups of entities with different media types:*
- *Respect user preferences by defining a hierarchy among media types.*
- *Incorporate the entity aligned with the user's preferred media type into the finalized summary.*
- *For groups of entities with the same media type:*
- *Evaluate the entity's role based on the partial order derived from rhetorical relations.*
- *Prioritize entities with more influential roles (e.g., main ideas, central concepts) in the finalized summary.*
- *If equal weights persist:*
- *Facilitate user intervention to select an entity.*
- *Empower users to make the final decision based on their preferences.*
- *Finalize the summary entities, considering the resolved cases of equal weighting.*
- *Return the finalized summary entities.*

This algorithm systematically addresses equal weighting situations, factoring in user preferences and the rhetorical roles of entities. By seamlessly incorporating user input and considering the structural significance of entities, the algorithm ensures the generation of an optimal summary that aligns with the user's preferences and encapsulates the document's key ideas.

*Example*: In the example of *The History of Space Exploration*, envision a scenario where two images exhibit equal weights. To resolve this, the algorithm leans on user media preferences. If the user favors images over text, the image entity is integrated into the finalized summary. Conversely, if the user prefers text, priority is given to the text entity. In parallel scenarios with two equally weighted sentences, their roles within rhetorical relations determine the prioritization. Should one sentence play a more central or pivotal role, it gains prominence in the finalized summary. Should equal weights persist, the algorithm defers to user intervention, allowing users to make decisions based on their preferences. This comprehensive approach guarantees that the resulting summary adheres to user preferences and effectively captures the document's essential content.

*H. Summary Presentation*

Upon finalizing the summarization process, which involves selecting, refining, and optimizing entities, the user is presented with the generated summary. This user-friendly and accessible presentation is customized to align with the original multimedia document's nature and the user's preferences, and can be delivered in various formats such as text, slides, or video clips. The summary presentation process includes these crucial steps:

1. *Entity Formatting* ensures smooth flow and readability by logically arranging the selected entities.
2. *Presentation Format Selection* chooses the most fitting summary format, considering the original document's characteristics and user preferences. Options include text-based summaries, slide presentations, or video clips.
3. *Device Adaptation* enhances accessibility and user experience by optimizing the presentation for various devices, including desktops, laptops, tablets, or mobile phones.
4. *Visual Element Integration*, when applicable, augments the presentation and provides a visual representation of the summarized content by integrating visual elements like images, graphs, or charts.

Through the systematic execution of these steps, the summary extraction process ensures the creation of concise and meaningful summaries from multimedia documents. These summaries are crafted to align with user preferences and the desired summary time.

*Example*: In the context of *The History of Space Exploration*, imagine the generated summary comprising a blend of text, images, and videos. The selected and trimmed entities are meticulously formatted to create a cohesive and coherent summary. The presentation format is customized based on user preferences, allowing for options like a text-based summary complemented by images or a multimedia presentation featuring slides and video clips. The presentation is optimized for diverse devices, ensuring an optimal viewing experience across desktops, laptops, tablets, and mobile phones. Incorporation of visual elements, such as images and videos, enhances the engagement and informativeness of the summary. This comprehensive presentation approach elevates accessibility and user experience, delivering a condensed yet enriching version of the original multimedia document.

*I. Implementation of the Multimedia Document Summarization Framework*

The following algorithm constitutes an implementation of the framework designed for document summarization. Grounded in a rhetorical relations-based model, the algorithm systematically navigates through various stages, considering user preferences and desired summary time. Our proposed algorithm orchestrates the summarization process through a series of well-defined functions and procedures, ensuring a tailored and coherent summary.

*1) Global Summarization Algorithm*

This algorithm orchestrates the entire process of generating a summary for a multimedia document. It begins by constructing a rhetorical graph to capture the document's structural relationships. Then, it weights and ranks the document entities based on their importance. Next, it trims the entities to fit within the specified summary duration while maintaining relevance. Finally, it resolves any cases where entities have equal weights by considering user preferences and rhetorical roles.

*Algorithm: GenerateSummary(doc, profile, duration)*
*graph = BuildGraph(doc)*
*entities = WeightAndRank(graph)*
*summary = Trim(entities, duration)*
*final_summary = ResolveEqual(summary, profile)*
*return final_summary*

*2) Constructing Rhetorical Graph*

This function creates a graph representation of the multimedia document's rhetorical structure. It identifies document units and their relationships

*Function BuildGraph(document)*
*graph = createEmptyGraph()*
*units = identifyUnits(document)*
*for each unit in units*
*addNode(graph, unit)*
*for each pair of units (unit1, unit2)*
*if unit1 != unit2*
*relation = determineRelation(unit1, unit2)*
*if relation exists*
*addEdge(graph, unit1, unit2, relation)*
*return graph*

*3) Weighting and Ranking Entities*

This function assigns weights to document entities based on their rhetorical importance and ranks them accordingly, providing a prioritized list of significant elements.

*Function WeightAndRank(graph)*
*weights = initializeWeights(graph)*
*traverseGraph(graph, weights)*
*ranked_entities = rankEntities(weights)*
*return ranked_entities*

*4) Trimming Entities for Time Constraints*

This function selects document entities that best fit within the specified summary duration while ensuring that the summary remains informative and concise.

*Function Trim(entities, summary_time)*
*trimmed_summary = []*
*current_time = 0*
*for each entity in entities*
*if current_time + entity.duration <= summary_time*
*append entity to trimmed_summary*
*current_time += entity.duration*
*else*
*break*
*return trimmed_summary*

*5) Resolving Cases of Equal Weights*

This function handles situations where entities possess equal weights, incorporating user preferences and rhetorical roles for optimal resolution.

*Function ResolveEqual(entities, profile)*
*final_summary = []*
*for each entity in entities*
*if entities.length == 1*
*append entity to final_summary*
*else*
*if entity.mediaType != previous_entity.mediaType*
*preferred_media = profile.preferredMedia ()*
*if entity.mediaType == preferred_media*
*append entity to final_summary*
*break*
*else*
*continue*
*else*
*role = determineRole(entity)*
*if role == 'MainIdea'*
*append entity to final_summary*
*break*
*if final_summary.isEmpty() and entities.isNotEmpty()*
*user_selected_entity = profile.selectEntity(entities)*
*append user_selected_entity to final_summary*
*return final_summary*

The global summarization algorithm provides anadaptive approach, offering a coherent and personalized summary. This systematic framework ensures flexibility and efficiency across diverse multimedia documents and user profiles, emphasizing the importance of considering both content and user-specific preferences. Please note that specific implementations of certain functions may vary based on system requirements.

## IV. ILLUSTRATIVE EXAMPLE

To provide a comprehensive illustration of the multimedia document summarization process, let's delve into the example of the document *The History of Space Exploration* which encompasses diverse units (Introduction, Space Race, Moon Landing, International Space Station, Mars Missions, and Future of Space Exploration, with the user expressing a preference for a 10-minute summary.

*Step 1: Document Specification*
- Multimedia Document: "The History of Space Exploration"
- User Profile: Preferred summary time = 10 minutes

Initiating the process, we set the stage by specifying the document and the user's preference, laying the groundwork for subsequent steps.

*Step 2: Rhetorical Relations Graph Construction*
The algorithm initiates by constructing a rhetorical relations graph that visually represents the hierarchical structure and relationships among different units in the document.
*Constructed graph*:
- Introduction -> Space Race
- Introduction -> Moon Landing
- Introduction -> International Space Station
- Space Race -> Moon Landing
- Moon Landing -> International Space Station
- International Space Station -> Mars Missions
- Mars Missions -> Future of Space Exploration

The graph highlights the rhetorical relationships between units, such as the connection of "Introduction" to "Space Race," "Moon Landing," and "International Space Station."

*Step 3: Weighting and Ranking of Entities*
Next, the algorithm assigns weights to each unit based on their significance within the document. These weights reflect the entities' relative importance and are utilized to rank them.

| Unit | Weight |
|---|---|
| Introduction | 0.65 |
| Space Race | 0.80 |
| Moon Landing | 0.90 |
| International Space Station | 0.85 |
| Mars Missions | 0.70 |
| Future of Space Exploration | 0.60 |

Weights are calculated considering factors such as entity relevance and user preferences, exemplified by the higher weight assigned to "Moon Landing" due to its crucial role in space exploration history.

*Step 4: Deletion of Least Important Entities*
In line with the 10-minute constraint, the algorithm trims entities, factoring in both weights and relevance. Suppose "Introduction" (3 minutes) and "Space Race" (4 minutes) are considered, and their combined duration fits the 10-minute limit. These entities, along with their rhetorical significance, make it to the trimmed summary.

*Step 5: Treatment of Cases of Equal Weighting*
Encountering equal weights (e.g., "Space Race" and "International Space Station"), the algorithm applies additional criteria. If the user favors space-related media, "International Space Station" gains priority. This adaptive approach ensures user-centric summarization.

*Step 6: Summary Presentation*
The final summary, comprising entities like "Introduction," "Moon Landing," and "International Space Station," aligns with user preferences. The format adapts to the original multimedia document, ensuring an engaging presentation encompassing text, images, and videos.

This example elucidates the systematic execution of our algorithm, demonstrating its adaptability to user preferences and constraints. Each step, from graph construction to summary presentation, follows the logical sequence defined by our algorithm, resulting in a concise and pertinent summary of the original multimedia document within the specified time.

## V. DISCUSSION & CONCLUSION

The proposed rhetorical relations-based framework for multimedia document summarization introduces innovative perspectives and avenues for further research and development. This section explores the implications, limitations, and potential future directions of the framework.

### A. Implications

The application of rhetorical relations in multimedia document summarization holds significant implications. By leveraging the inherent structure and coherence of documents, the framework facilitates the extraction of key information, resulting in concise and coherent summaries. This approach enhances users' efficiency in navigating extensive multimedia content, saving time while achieving a comprehensive understanding of the document.

Moreover, the utilization of weighting algorithms enables the prioritization of units based on importance, ensuring that summaries capture the most relevant aspects. This not only enhances the readability of summaries but also assists users in focusing on crucial information, thereby improving comprehension and decision-making.

Furthermore, the framework's versatility in handling diverse media types, including text, images, and videos, broadens its applicability to various multimedia sources. This adaptability allows for more comprehensive and inclusive summarization across a wide range of content, contributing to its practical utility in diverse domains such as journalism, education, and content curation.

## B. Limitations and Challenges

While the framework shows promise, it is essential to acknowledge its limitations and challenges. Accurately determining the weights and durations of multimedia units poses a primary challenge, necessitating the development of robust algorithms capable of effectively assessing the relevance and importance of different units across various media formats.

Additionally, scalability presents another challenge, particularly as multimedia documents increase in size and complexity. Optimizing algorithms and exploring techniques such as parallel computing are crucial for maintaining efficiency during the summarization process, especially when dealing with large-scale datasets and real-time summarization requirements.

Moreover, the subjectivity of user preferences presents a challenge in personalizing summaries. While the framework considers user profiles, refining the process to determine the most suitable units requires a deeper understanding of individual preferences. Exploring advanced user modeling techniques and incorporating user feedback mechanisms can help address this challenge and improve the overall personalization effectiveness.

## C. Future Directions

The proposed framework suggests avenues for future research and development. Advanced weighting algorithms, integrating machine learning techniques and contextual information, could enhance unit importance assessment precision. Leveraging data analytics and NLP methods could refine the framework's ability to prioritize key content elements across multimedia sources. Moreover, integrating NLP techniques like entity recognition, sentiment analysis, and topic modeling could improve summary quality. Incorporating semantic analysis and context-aware processing could lead to more informative summaries.

Expanding the framework to include multimodal fusion techniques is recommended for comprehensive summarization. Indeed, leveraging advances in computer vision, audio processing, and cross-modal understanding can lead to richer summaries. Additionally, focusing on user interaction and evaluation methods is crucial. Conducting user studies and gathering feedback are essential for enhancing usability and effectiveness. By adopting a user-centered design approach, researchers can ensure that the framework meets diverse user needs.

## D. Conclusion

In conclusion, the rhetorical relations-based framework for multimedia document summarization offers a valuable approach for extracting key information and constructing concise, coherent summaries. By leveraging rhetorical relations, weighting algorithms, and user preferences, the framework enables personalized and contextually relevant summarization experiences. Despite challenges and limitations, the framework presents significant implications and exciting opportunities for future research. By addressing identified challenges, exploring advanced techniques, and incorporating user feedback, researchers can further refine the framework, contributing to more effective and efficient multimedia document summarization. The systematic execution of different steps results in a concise and meaningful summary tailored to user preferences and desired summary time, underscoring its value in efficient summarization.